\documentclass[journal]{IEEEtran}
\usepackage{booktabs}

\usepackage{cite}
\ifCLASSINFOpdf
   \usepackage[pdftex]{graphicx}
\else
 \usepackage[dvips]{graphicx}
\fi

\usepackage[cmex10]{amsmath}
\usepackage{amssymb}

\usepackage{url}

\begin{document}

%
\title{Brain-Computer Interface Controlled Robotic Gait Orthosis}
%
%
%

\author{An H. Do$^{1}$\IEEEmembership{,}
        Po T. Wang$^{2}$\IEEEmembership{,}
                                Christine E. King$^{2}$\IEEEmembership{,}
                                Sophia N. Chun$^{3}$\IEEEmembership{,}
        Zoran Nenadic$^{2,4}$\IEEEmembership{}

\thanks{$^{1}$Department of Neurology, University of California, Irvine, CA, USA, email: and@uci.edu}%
\thanks{$^{2}$Department of Biomedical Engineering, UCI, CA, USA}%
\thanks{$^{3}$Department of Spinal Cord Injury, Long Beach Veterans Affairs Medical Center, Long Beach, CA, USA}%
\thanks{$^{4}$Department of Electrical Engineering and Computer Science, UCI, CA, USA}}%

\maketitle

\begin{abstract}
\textbf{Background}: Excessive reliance on wheelchairs in individuals with tetraplegia or paraplegia due to spinal cord injury (SCI) leads to many medical co-morbidities, such as cardiovascular disease, metabolic derangements, osteoporosis, and pressure ulcers. Treatment of these conditions contributes to the majority of SCI health care costs. Restoring able-body-like ambulation in this patient population can potentially reduce the incidence of these medical co-morbidities, in addition to increasing independence and quality of life. However, no biomedical solution exists that can reverse this loss of neurological function, and hence novel methods are needed. Brain-computer interface (BCI) controlled lower extremity prostheses may constitute one such novel approach.

\textbf{Methods}: One able-bodied subject and one subject with paraplegia due to SCI underwent electroencephalogram (EEG) recordings while engaged in alternating epochs of idling and walking kinesthetic motor imagery (KMI). These data were analyzed to generate an EEG prediction model for online BCI operation. A commercial robotic gait orthosis (RoGO) system (suspended over a treadmill) was interfaced with the BCI computer to allow for computerized control. The subjects were then tasked to perform five, 5-min-long online sessions where they ambulated using the BCI-RoGO system as prompted by computerized cues. The performance of this system was assessed with cross-correlation analysis, and omission and false alarm rates.

\textbf{Results}: The offline accuracy of the EEG prediction model averaged 86.30\% across both subjects (chance: 50\%). The cross-correlation between instructional cues and the BCI-RoGO walking epochs averaged across all subjects and all sessions was 0.812$\pm$0.048 (p-value$<$$10^{-4}$). Also, there were on average 0.8 false alarms per session and no omissions. 

\textbf{Conclusion}: These results provide preliminary evidence that restoring brain-controlled ambulation after SCI is feasible. Future work will test the function of this system in a population of subjects with SCI. If successful, this may justify the future development of BCI-controlled lower extremity prostheses for free overground walking for those with complete motor SCI. Finally, this system can also be applied to incomplete motor SCI, where it could lead to improved neurological outcomes beyond those of standard physiotherapy. 

\end{abstract}


 \ifCLASSOPTIONpeerreview
 \begin{center} \bfseries EDICS Category: 3-BBND \end{center}
 \fi
%
\IEEEpeerreviewmaketitle

\section{Introduction}

\IEEEPARstart{I}ndividuals with tetraplegia or paraplegia due to spinal cord injury (SCI) are unable to walk and most are wheelchair bound. Decreased physical activity associated with prolonged wheelchair use leads to a wide range of co-morbidities such as metabolic derangements, heart disease, osteoporosis, and pressure ulcers~\cite{rljohnson:96}.  Unfortunately, no biomedical solutions can reverse this loss of neurological function, and treatment of these co-morbidities contributes to the bulk of medical care costs for this patient population~\cite{rljohnson:96}. While commercially available lower extremity prostheses can help restore basic ambulation via robust manual control, their adoption among the SCI community remains low, likely due to cost, bulkiness, and energy expenditure inefficiencies. 
Hence, novel approaches are needed to restore able-bodied-like ambulation in people with SCI. If successful, these will improve the quality of life in this population, and reduce the incidence and cost of medical co-morbidities as well as care-giver burden. 

A brain-computer interface (BCI) controlled lower extremity prosthesis may be one such novel approach. It can be envisioned that a combination of an invasive brain signal acquisition system and implantable functional electrical stimulation (FES) electrodes can potentially act as a permanent BCI prosthesis. However, for safety reasons, the feasibility of brain-controlled ambulation must first be established using noninvasive systems.  

This concept was explored in the authors' prior work~\cite{ptwang:10,ptwang:12,ceking:13} in which subjects (both able-bodied and SCI) used an electroencephalogram (EEG) based BCI to control the ambulation of an avatar within a virtual reality environment. In these studies, subjects utilized idling and walking kinesthetic motor imagery (KMI) to complete a goal-oriented task of walking the avatar along a linear path and making stops at 10 designated points. In addition, two out of five subjects with SCI achieved BCI control that approached that of a manually controlled joystick. While these results suggest that accurate BCI control of ambulation is possible after SCI, the translation of this technology from virtual reality to a physical prosthesis has not been achieved. In this study, the authors report on the first case of integrating an EEG-based BCI system with a robotic gait orthosis (RoGO), and its successful operation by both able-bodied and SCI subjects.

\section{Methods}
To facilitate the development of a BCI-controlled RoGO, EEG data were recorded from subjects as they engaged in alternating epochs of idling and walking KMI. These data were then analyzed offline to generate an EEG prediction model for online BCI operation. A commercial RoGO system (suspended over a treadmill), was interfaced with the BCI computer to allow for computerized control. In a series of five, 5-min-long online tests, the subjects were tasked to ambulate using the BCI-RoGO system when prompted by computerized cues. The performance of this system was assessed by calculating the cross-correlation and latency between the computerized cues and BCI-RoGO response, as well as the omission and false alarm rates. 

\subsection{Training Data Acquisition}\label{sec:satda}
Ethical approval was obtained from the Institutional Review Board at the Long Beach Veterans Affairs Medical Center (LBVA) and the University of California, Irvine (UCI). Subjects were recruited from a population of able-bodied individuals, or those with chronic, complete motor paraplegia due to SCI ($>$ 12 months post-injury). The exclusion criteria for subjects with SCI were severe spasticity, contractures, restricted range of motion, or fractures in the lower extremities, pressure ulcers, severe osteoperosis, or orthostatic hypotension. These criteria were ruled out in a safety screening evaluation consisting of an interview, a physical exam, lower extremity dual-energy x-ray absorptiometry (DEXA) scan and x-rays, and a tilt-table exam.

\begin{figure}
\centering
\includegraphics[scale= 0.4]{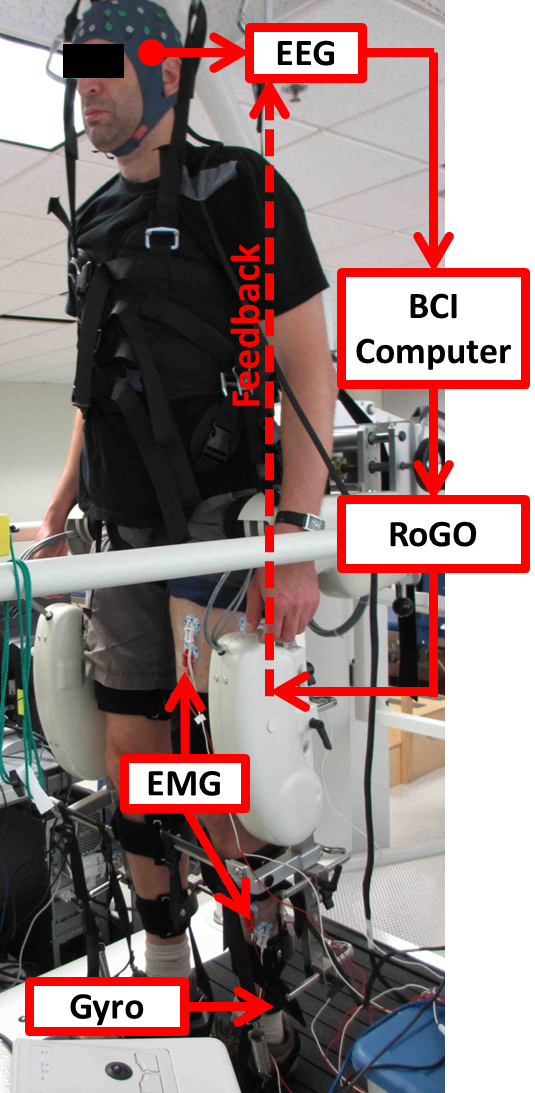}
\caption{The experimental setup showing a subject suspended in the RoGO, while donning an EEG cap, surface EMG electrodes, and a gyroscope on the left leg. A monitor (not shown), placed in front of the subject at eye-level, presented instructional cues. }
\label{fig:Figure1}
\end{figure}

An actively shielded 64-channel EEG cap was first mounted on the subjects' head and impedances were reduced to $<$10 K$\Omega$. EEG signals were acquired using 2 linked NeXus-32 bioamplifiers (Mind Media, Roermond-Herten, The Netherlands) at a sampling rate of 256 Hz. The subjects were suspended into a treadmill-equipped RoGO (Lokomat, Hocoma, Volketswil, Switzerland) using partial weight unloading (see Figure \ref{fig:Figure1}). Note that unlike overground orthoses, this system facilitates safe and easy testing conditions for the early development of BCI-prostheses for ambulation. Finally, EEG data were collected as the subjects alternated between 30-sec epochs of idling and walking KMI for a total of 10 min, as directed by computer cues. This entails vivid imagination of walking during walking KMI cues, and relaxation during idling cues. During this procedure,  the subjects stood still with their arms at their sides. 

\subsection{Electromyogram and Leg Movement Measurement}\label{sec:EMG}
Electromyogram (EMG) was measured to rule out BCI control by voluntary leg movements in able-bodied subjects. To this end, baseline lower extremity EMG were measured under 3 conditions: \textit{active walking} (subject voluntarily walks while the RoGO servos are turned off); \textit{cooperative walking} (subject walks synergistically with the RoGO); and \textit{passive walking} (the subject is fully relaxed while the RoGO makes walking movements). Three pairs of surface EMG electrodes were placed over the left quadriceps, tibialis anterior, and gastrocnemius (Figure \ref{fig:Figure1}), and signals were acquired with a bioamplifier (MP150, Biopac, Goleta, CA), bandpass filtered (0.1-1000 Hz), and sampled at 4 KHz. In addition, leg movements were measured by a gyroscope (Wii Motion Plus, Nintendo, Kyoto, Japan) with a custom wristwatch-like enclosure, strapped to the distal left lower leg (proximal to the ankle, see Figure \ref{fig:Figure1})~\cite{kaminian:02}. Approximately 85\% body-weight unloading was necessary for proper RoGO operation. The walking velocity was set to 2 km/hr. 

\subsection{Offline Analysis}\label{sec:ofa}
An EEG prediction model was generated using the methods described in Wang \textit{et al.}~\cite{ptwang:12}, which are briefly summarized here. First, the training EEG data were subjected to an automated algorithm to exclude those EEG channels with excessive artifacts. The EEG epochs corresponding to ``idling'' and ``walking'' states were then transformed into the frequency domain, and their power spectral densities (PSD) were integrated over 2-Hz bins. The data then underwent dimensionality reduction using a combination of classwise principal component analysis (CPCA)~\cite{kdas:07,kdas:09} and approximate information discriminant analysis (AIDA)~\cite{Das2008}. The resulting 1-D spatio-spectral features were extracted by:
\begin{equation}
\label{eq:feature}
f = T_A\Phi_C(d)
\end{equation}
where $f\in\mathbb{R}$ is the feature, $d\in \mathbb{R}^{B\times C}$ are single-trial spatio-spectral EEG data ($B$-the number of frequency bins, $C$-the number of retained EEG channels), $\Phi_C:\mathbb{R}^{B\times C}\rightarrow \mathbb{R}^m$ is a piecewise linear mapping from the data space to the $m$-dimensional CPCA-subspace, and $T_A: \mathbb{R}^{m} \rightarrow \mathbb{R}$ is an AIDA transformation matrix. Detailed descriptions of these techniques are found in~\cite{kdas:09, Das2008}. A linear Bayesian classifier: 

\begin{equation}
\label{eq:lrt}
f^{\star} \in
\begin{cases}
\mathcal{I}, &\text{if}\quad P(\mathcal{I}\,|f^{\star})>P(\mathcal{W}\,|f^{\star}) \\
\mathcal{W}, &\text{otherwise} 
\end{cases}
\end{equation}
was then designed in the feature domain, where $P(\mathcal{I}\,|f^{\star})$ and $P(\mathcal{W}\,|f^{\star})$ are the posterior probabilities of ``idling'' and ``walking'' classes, respectively. 
The performance of the Bayesian classifier~(\ref{eq:lrt}), expressed as a classification accuracy, was then assessed by performing stratified 10-fold cross-validation~\cite{rkohavi:95}. This was achieved by using 90\% of the EEG data to train the parameters of the CPCA-AIDA transformation and the classifier. The remaining 10\% of the data then underwent the above transformation and classification. This process was repeated 10 times, each time using a different set of 9 folds for training and the remaining 1 fold for testing. Finally, the optimal frequency range $[F_L, F_H]$ was found by increasing the lower and upper frequency bounds and repeating the above procedure until the classifier performance stopped improving~\cite{ahdo:11}. The parameters of the prediction model, including $[F_L, F_H]$, the CPCA-AIDA transformation, and the classifier parameters, were then saved for real-time EEG analysis during online BCI-RoGO operation. The above signal processing and pattern recognition algorithms were implemented in the BCI software and were optimized for real-time operation~\cite{ahdo:11}. 

\subsection{BCI-RoGO Integration}
To comply with the institutional restrictions that prohibit software installation, the RoGO computer was interfaced with the BCI using a pair of microcontrollers (Arduino, SmartProjects, Turin, Italy) to perform mouse hardware emulation. Microcontroller \#1 relayed commands from the BCI computer to microcontroller \#2 via an Inter-Integrated Circuit (I$^{2}$C) connection. Microcontroller \#2 then acted as a slave device programmed with mouse emulation firmware \cite{darran:11} to automatically manipulate the RoGO's user interface. This setup enabled the BCI computer to directly control the RoGO idling and walking functions. 

\subsection{Online Signal Analysis}

During online BCI-RoGO operation, 0.75-sec segments of EEG data were acquired every 0.25 sec in a sliding overlapping window. The PSD of the retained EEG channels were calculated for each of these segments and used as the input for the signal processing algorithms described in Section~\ref{sec:ofa}. The posterior probabilities of idling and walking classes were calculated using the Bayes rule, as explained in Section \ref{sec:ofa}. 

\subsection{Calibration}\label{sec:calibration}
Similar to the authors' prior work~\cite{ptwang:10, ahdo:11, ptwang:12, ceking:11, ahdo:12, ceking:13}, the BCI-RoGO system is modeled as a binary state machine with ``idling'' and ``walking'' states. 
This step is necessary to reduce noise in the online BCI operation and minimize the mental workload of the subject. To this end, the posterior probability was averaged over 2 sec of EEG data, $\bar{P}(W|f^{\star})$, and compared to two thresholds, $T_{I}$ and $T_W$, to initiate state transitions. Specifically, the system transitioned from ``idling'' to ``walking''  state (and vice versa) when $\bar{P}>T_{W}$ ($\bar{P}<T_{I}$), respectively. Otherwise, the system remained in the current state. 

The values of $T_{I}$ and $T_{W}$ were determined from a short calibration procedure. Specifically, the system was set to run in the online mode (with the RoGO walking disabled) as the subject alternated between idling or walking KMI for $\sim$5 min. The values of $\bar{P}$ were plotted in a histogram to empirically determine the values of $T_{I}$ and $T_{W}$. A brief familiarization online session with feedback was used to further fine-tune these threshold values. 

\subsection{Online Evaluation}\label{sec:oe}
In an online evaluation, the subjects, while mounted in the RoGO, used idling/walking KMI to elicit 5 alternating 1-min epochs of BCI-RoGO idling/walking, as directed by static, textual computer cues (see Additional Files for videos). Ideally, during walking KMI, the underlying EEG changes should initiate and maintain BCI-RoGO walking until walking KMI stops. The subjects were instructed to make no voluntary movements and to keep their arms still at their side. Left leg EMG and movements were measured as described in Section~\ref{sec:EMG}. This online test was performed 5 times in a single experimental day. 

Online performance was assessed using the following metrics~\cite{ceking:11,ahdo:11,ahdo:12}: 
\begin{enumerate}
	\item Cross-correlation between the cues and BCI-RoGO walking 
	\item Omissions (OM)---failure to activate BCI-RoGO walking during the ``Walk'' cues
	\item False Alarms (FA)---initiation of BCI-RoGO walking during the ``Idle'' cues  
\end{enumerate}

For able-bodied subjects, analysis of EMG and leg movement data was performed to ascertain whether RoGO walking was entirely BCI controlled. First, to demonstrate that covert movements were not used to initiate BCI-RoGO walking, gyroscope and rectified EMG data (in the 40-400 Hz band) were compared to the BCI decoded ``walking''  states in each session. Ideally, the initiation of these states should precede EMG activity and leg movements. Then, to establish whether voluntary movements were used to maintain BCI-RoGO walking, EMG during these epochs were compared to the baselines (see Section~\ref{sec:EMG}). To this end, EMG data were segmented by individual steps based on the leg movement pattern~\cite{kaminian:02}, as measured by the gyroscope. The PSD of these EMG segments were then averaged and compared to those of the baseline walking conditions. Ideally, the EMG power during BCI-RoGO walking should be similar to that of \textit{passive walking} and different from those of \textit{active} and \textit{cooperative walking}.

\subsection{Controls}
To determine the significance of each online BCI-RoGO session's performance, a nonlinear auto-regressive model was created:
\begin{eqnarray}
\label{eq:ar}
X_{k+1} &=& \alpha X_k + \beta W_k,\quad X_0\sim\mathcal{U}(0,1) \\
Y_k &=& h(X_k)
\end{eqnarray} 
where $X_{k}$ is the state variable at time $k$, $W_{k}\sim\mathcal{U}\left(0,1\right)$ is uniform white noise, $Y_k$ is the simulated posterior probability, and $h$ is a saturation function that ensures $Y_k\in\left[0,1\right]$. Let $P_k:=P(\mathcal{W}|f_k^{\star})$ be a sequence of online posterior probabilities calculated in Section \ref{sec:ofa}. Assuming the sequence $\{P_k\}$ is wide-sense stationary with mean $\mu$ and variance $\sigma^2$, the coefficients $\alpha$ and $\beta$ can be determined from:   
\begin{eqnarray}
\label{eq:}
\alpha &=& \rho\\
\alpha \mu + \frac{\beta}{2}&=&\mu
\end{eqnarray}      
where $\rho$ is the correlation coefficient between $P_{k+1}$ and $P_k$. Using these coefficients, 10,000 Monte Carlo trials were performed for each online session. Each sequence of simulated posteriors, $\{Y_k\}$, was then processed as in Section \ref{sec:calibration} above, and the cross-correlation between the cues and simulated BCI-RoGO state sequence was calculated. An empirical p-value was defined as a fraction of Monte Carlo trials whose maximum correlation was higher than that of the online session.

\section{Results}

\begin{figure}
\centering
\includegraphics[width=\linewidth]{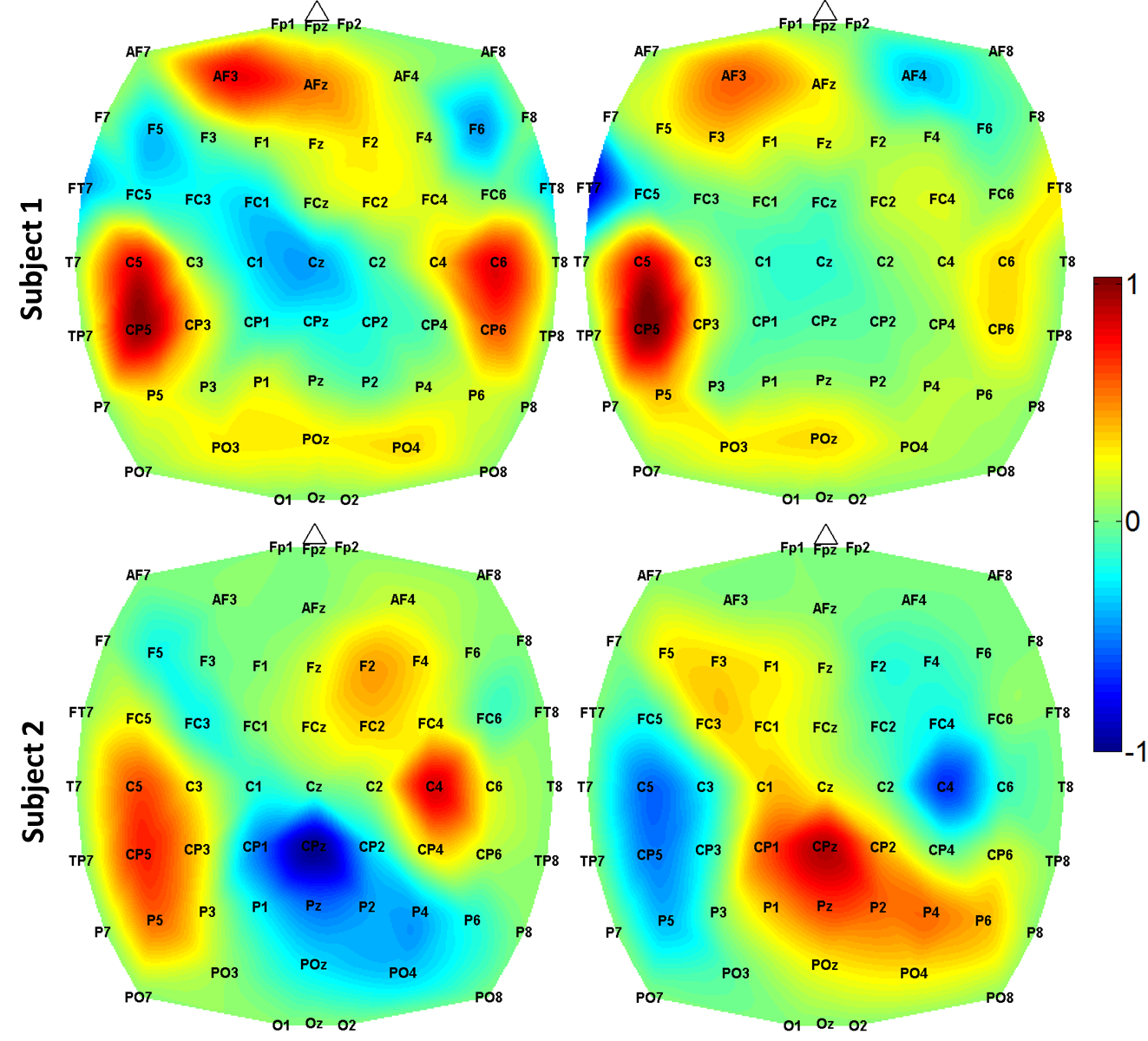}
\caption{The CPCA-AIDA feature extraction maps for both subjects.   Since feature extraction is piecewise linear, there is one map for each of the 2 classes. Brain areas with values close to +1 or -1 are most salient for distinguishing between idling and walking classes at this frequency. The most salient features were in the 8-10 Hz bin for Subject 1 and the 10-12 Hz bin for Subject 2.}
\label{fig:Figure2}
\end{figure}

Two subjects (one able-bodied and one with paraplegia due to SCI) were recruited for this study and provided their informed consent to participate. Their demographic data are described in Table 1 below. Subject 2, who was affected by paraplegia due to SCI, underwent the screening evaluation and met all study criteria. All subjects successfully underwent the training EEG procedure. Their EEG prediction models were generated as described in Section~\ref{sec:ofa} based on training EEG data (Section~\ref{sec:satda}). This offline analysis resulted in a model classification accuracy of 94.8$\pm$0.8\% and 77.8$\pm$2.0\% for Subjects 1 and 2, respectively (chance: 50\%). The EEG feature extraction maps are shown in Figure \ref{fig:Figure2}. After the calibration procedure (Section~\ref{sec:calibration}), a histogram of posterior probabilities was plotted (Figure 3). Based on this histogram and a familiarization trial, the respective values of $T_I$ and $T_W$ were set at 0.04 and 0.65 for Subject 1, and 0.50 and 0.90 for Subject 2. 

\begin{table}
\centering
\caption{Demographic data of the study subjects. ASIA =  American Spinal Injury Association.}
  \begin{tabular}{|c|c|c|c|c|}
  	\hline
  	Subject & Age & Gender & Prior BCI Experience & SCI Status  \\ \hline
  	  1    & 42  & Male &  $\sim$5 hours     & N/A \\ \hline
  	  2   & 25  &  Male & $\sim$3 hours     & T6 ASIA B   \\ \hline
  \end{tabular} 
\end{table}

\begin{figure}[ht]
\centering
\includegraphics[width=\linewidth]{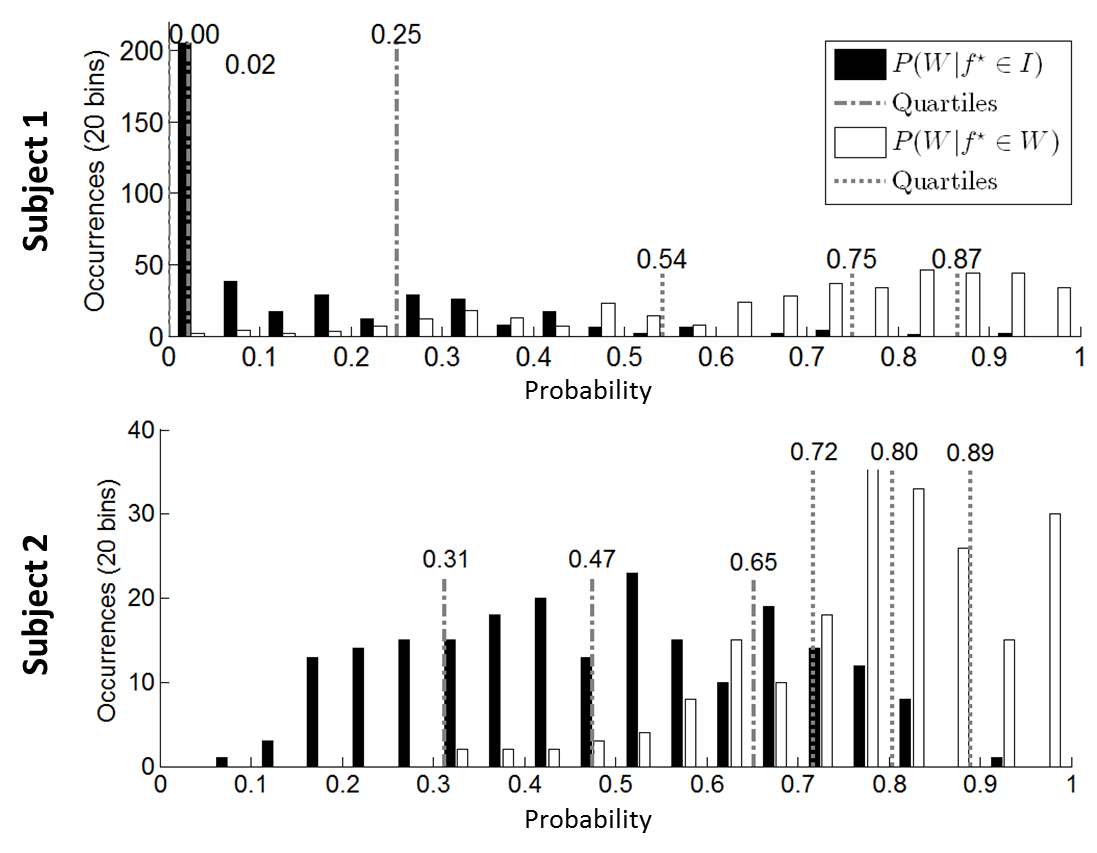}
\caption{A representative histogram of averaged posterior probabilities, $\bar{P}(W|f^{\star})$ for both subjects.}
\label{fig:Figure3}
\end{figure}

The performances from the 5 online sessions for both subjects are summarized in Table 2. The average cross-correlation between instructional cues and the subjects' BCI-RoGO walking epochs was 0.812$\pm$0.048. As a control, the maximum cross-correlation between the instructional cues and simulated BCI operation using 10,000 Monte Carlo trials were 0.438 and 0.498 for Subjects 1 and 2, respectively. This indicates that all of the cross-correlations in Table 2 were significant with an empirical p-value $<$ $10^{-4}$. Also, there were no omissions for either subject. The false alarm rate averaged 0.8 across all sessions and both subjects. While the duration of these false alarm epochs averaged 7.42$\pm$2.85 sec, much of this time can be attributed to the RoGO's locked-in startup sequence ($\sim$5 sec). In addition, each subject managed to achieve 2 sessions with no false alarms. Videos of a representative online session for the able bodied subject (Subject 1) and for the subject with SCI (Subject 2) are provided as downloadable supplemental media. Alternatively, online versions can be found at: http://www.youtube.com/watch?v=W97Z8fEAQ7g, and http://www.youtube.com/watch?v=HXNCwonhjG8.

\begin{figure}
\centering
\includegraphics[width=\linewidth]{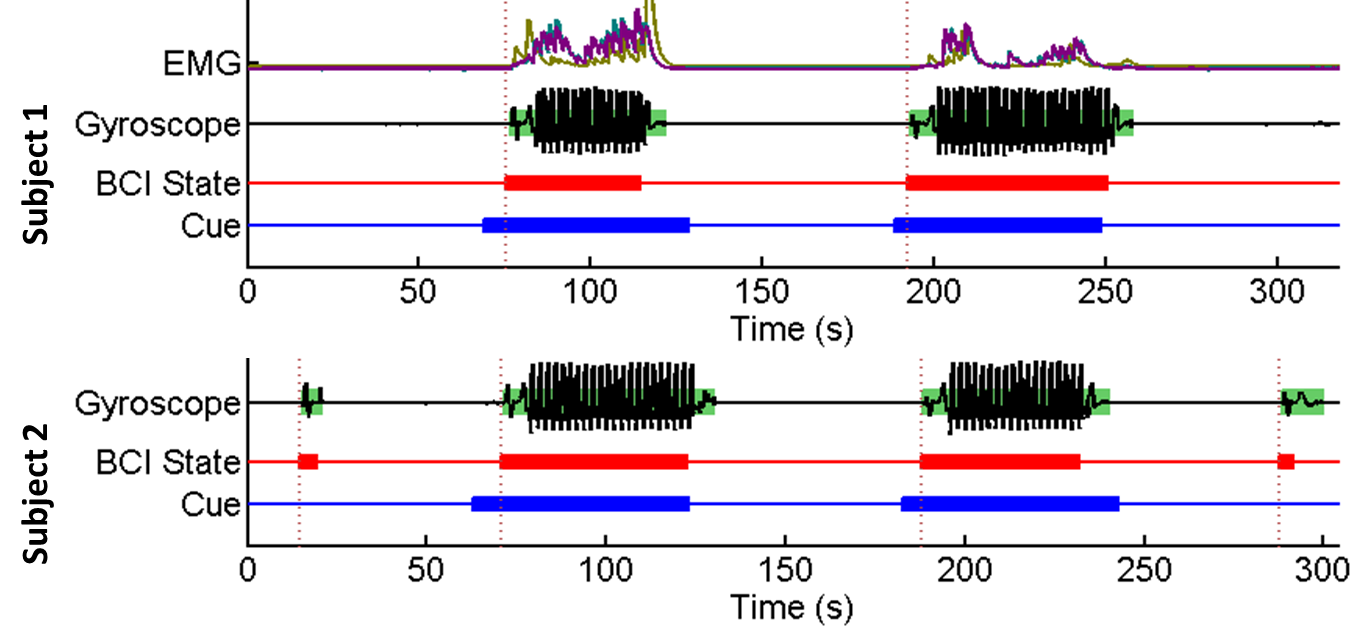}
\caption{Results from a representative online session for each subject, showing epochs of idling and BCI-RoGO walking determined from the gyroscope trace (green blocks). The red trace represents the decoded BCI states, while the blue trace represents the instructional cues. The thick/thin blocks indicate walking/idling. Corresponding EMG (gold: quadriceps; teal: tibialis anterior; purple: gastrocnemius) are also shown. Note that EMG was not measured for Subject 2.}
\label{fig:Figure4}
\end{figure}

\begin{figure}
\centering
\includegraphics[width=\linewidth]{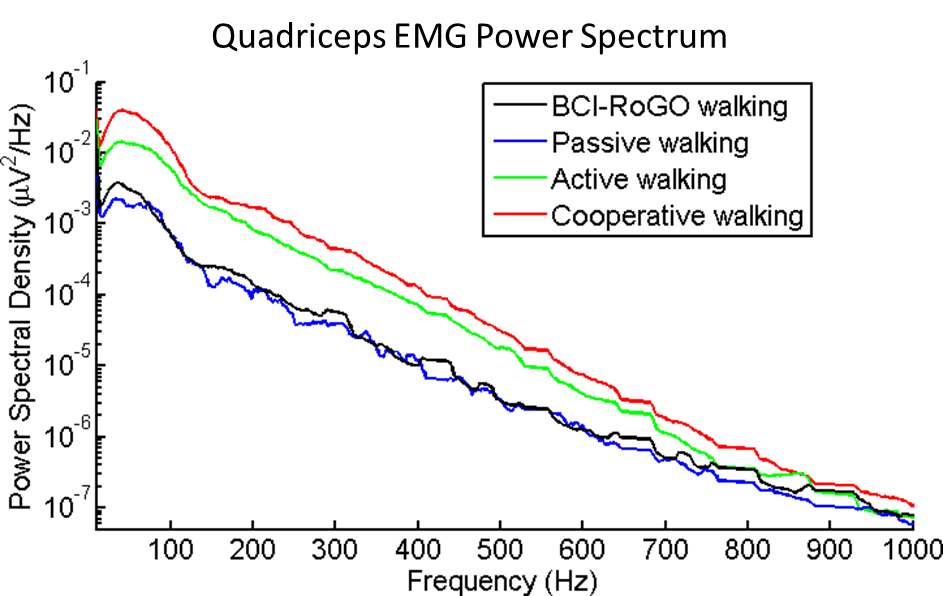}
\caption{Representative EMG PSD of the quadriceps for Subject 1, demonstrating that EMG during BCI-RoGO walking are different from active or cooperative walking baseline conditions, and are similar to passive walking.}
\label{fig:Figure5}
\end{figure}

For Subject 1, the EMG and leg movement data from online sessions were analyzed as described in Section~\ref{sec:oe}. EMG and gyroscope measurements indicated that no movement occurred prior to the initiation of BCI decoded ``walking'' states (see Figure \ref{fig:Figure4}). When compared to the baselines, the EMG during online BCI-RoGO walking in all 3 muscle groups were statistically different from those of \textit{active} or \textit{cooperative walking} conditions (p $<$ 10$^{-13}$), and were not different from those of passive walking (p = $0.37$). These results confirm that the BCI-RoGO system was wholly BCI controlled. Note that \textit{passive walking} is known to generate EMG activity~\cite{smazzoleni:09}, hence a similar level of activity during BCI-RoGO walking (Figure \ref{fig:Figure5}) is expected. Furthermore, since Subject 2 does not have any voluntary motor control of the lower extremities, there was no need to perform EMG measurements and analysis.

\begin{table*}
\centering
\caption{Cross-correlation between the BCI-RoGO walking and cues at specific lags, number of omissions and false alarms, and the average duration of false alarm epochs. }
\begin{tabular}{|c|c|c|c|c|}
\hline
 & Session & Cross-correlation (lag in sec) & Omissions & False Alarms (avg. duration in sec)               \\  \hline 
Subject 1 & 1  &      0.771   (10.25) &    0 & 1 (12.00)   \\  \hline
& 2  & 0.741 (4.50)         &    0 & 2 (5.50$\pm$0.00)\\  \hline
& 3  & 0.804 (3.50)         &    0 & 1 (5.30)  \\  \hline
& 4  & 0.861 (4.50)         &    0 & 0        \\  \hline
& 5  & 0.870 (12.00)        &    0 & 0        \\  \hline
& Avg.& 0.809$\pm$0.056 (6.95$\pm$3.89) & 0  & 0.8 (7.08$\pm$3.28)                           \\ \hline 
Subject 2 & 1 & 0.781 (6.25) & 0 & 1 (8.80) \\ \hline
& 2 & 0.878 (6.75) & 0 & 0 \\ \hline
& 3 & 0.782 (6.25) & 0 & 0 \\ \hline
& 4 & 0.851 (14.25) & 0  & 1 (5.50) \\ \hline
& 5 & 0.785 (5.75) & 0 & 2 (8.40$\pm$4.10) \\ \hline
& Avg. & 0.815$\pm$0.046 (7.85$\pm$3.60)  & 0 & 0.8 (7.76$\pm$2.80) \\ \hline
& Overall Avg. & 0.812$\pm$0.048 (7.40$\pm$3.56) & 0 & 0.8 (7.42$\pm$2.85) \\ \hline
\end{tabular}
\end{table*}

\section{Discussion and Conclusion}
The results of this study demonstrate that BCI-controlled lower extremity prostheses for walking are feasible. Both subjects gained purposeful and highly accurate control of the BCI-RoGO system on their first attempt. It is particularly notable that the subject with paraplegia due to SCI (Subject 2) was able to accomplish this with minimal prior BCI experience and after only a brief 10 min training data acquisition session. To the best of the authors' knowledge, this represents the first-ever demonstration of a person with paraplegia due to SCI re-gaining brain-driven basic ambulation and completing a goal-oriented walking task. 

The EEG prediction models for both subjects in this study had a high offline classification accuracy. In the case of Subject 1, the performance was higher than his performances in prior BCI walking avatar studies (Subject 1 in \cite{ptwang:10}, and Subject A1 in \cite{ptwang:12}). Note that the gain in performance was achieved despite the subject being suspended in the RoGO (as opposed to being seated as in~\cite{ptwang:10,ptwang:12}). Examination of the prediction models also revealed that the salient brain areas underlying walking KMI varied from subject to subject (Figure \ref{fig:Figure2}). Collectively, these areas likely overlie the pre-frontal cortex, supplementary motor, and the leg and arm sensorimotor representation areas, and are consistent with those previously reported. For example, activation of the pre-frontal cortex and supplementary motor area during walking motor imagery has been described in functional imaging studies \cite{clafougere:10}. Similarly, involvement of the leg and bilateral arm areas during walking KMI have been reported in~\cite{ptwang:12,ceking:13} and may be associated with leg movement and arm swing imagery. This EEG prediction model was further validated by generating highly separable posterior probability distributions (Figure \ref{fig:Figure3}) and facilitating highly accurate online BCI-RoGO control. Finally, since it was generated through a data-driven procedure, the modeling approach is subject specific and may accommodate for the neurophysiological variability across subjects~\cite{ptwang:10,ptwang:12,ceking:13}.   

Both subjects attained highly accurate online control of the BCI-RoGO system. This was achieved immediately on each subjects' first attempt and generally improved through the course of the 5 online sessions. The average online cross-correlation between the computer cues and BCI response (0.812) was higher than those achieved with lower (0.67) and upper (0.78) extremity BCI-prostheses~\cite{ahdo:11,ceking:11}, despite EEG being acquired under more hostile (ambulatory) conditions. Furthermore, not only did the subject with paraplegia attain immediate BCI-RoGO control, but he also had a higher average online performance than the able-bodied subject. This implies that future BCI-prostheses for restoring overground walking after SCI may be feasible. Additionally, all of the subjects' online BCI-RoGO sessions were purposeful with a 100\% response rate (no omissions). Although Subject 1 had no false alarms by the end of the experiment, Subject~2  still experienced false alarms in the final session. Although few in number and short in duration, false alarms carry the risk of bodily harm, and this problem must be addressed in the development of future BCI-prostheses for overground walking. Table 2 also shows that the maximum correlation is attained at an average lag of 7.4 sec. Most of this lag can be attributed to the RoGO's locked-in power-down sequence ($\sim$5 sec). Minor sources of delay include a combination of user response time and the 2-sec long posterior probability averaging window (see Section~\ref{sec:calibration}). This delay can potentially be minimized with additional user training in a controlled environment. Also,  reducing the averaging window may eliminate some of the delay, but this would be at the expense of increasing the false alarm and omission rates. This trade-off will be examined in future studies.  

With no more than $\sim$5 hr of relevant BCI experience (operating the BCI walking avatar as described in \cite{ptwang:10,ptwang:12,ceking:13}), both subjects attained a high-level of control of the BCI-RoGO system after undergoing a series of short procedures (i.e. 10 min training data acquisition, 5 min calibration, 5 min familiarization). This indicates that a data-driven EEG prediction model as well as prior virtual reality training may have facilitated this rapid acquisition of BCI control. In addition, this model enables BCI operation using an intuitive control strategy, i.e. walking KMI to induce walking and idling KMI to stop. This is in contrast to requiring subjects to undergo months of training in order to acquire a completely new skill of modulating pre-selected EEG signal features as frequently done in operant conditioning BCI studies. However, it remains unclear whether applying an EEG decoding model generated from idling/walking KMI will be robust enough against EEG perturbations caused by other simultaneous cognitive and behavioral processes common during ambulation (e.g. talking, head turning). Anecdotally, no disruption of BCI operation was observed in this study and related previous BCI studies \cite{ptwang:12,ceking:13} when the subjects engaged in brief conversations or hand and arm gestures during the familiarization session. Formalized testing of this hypothesis would require additional studies to be performed.

Based on the above observations, this data-driven BCI approach may be necessary for future intuitive and practical BCI-controlled lower extremity prostheses for people with SCI. This approach would enable subjects with SCI to use intuitive BCI control strategies such as KMI of walking or attempted (albeit ineffective) walking. Similar to Subject 2 in this study, this can potentially be accomplished with minimal user training and supervision from the experiment operator. Finally, this approach may enhance the appeal and practicality of future BCI-controlled lower extremity prostheses for ambulation by reducing the time burden and associated costs. 

In conclusion, these results provide convincing evidence that BCI control of ambulation after SCI is possible, which warrants future studies to test the function of this system in a population of subjects with SCI. Since participants with SCI were able to operate the BCI-walking simulator~\cite{ptwang:12,ceking:13}, it is expected that they can readily transfer these skills to the BCI-RoGO system, similar to Subject 2. If successful, such a system may justify the future development of BCI-controlled lower extremity prostheses for free overground walking for those with complete motor SCI. This includes addressing issues such as additional degrees of freedom (e.g. turning, velocity modulation, transitioning between sitting and standing), as well as appropriate solutions for signal acquisition (e.g. invasive recordings). Finally, the current BCI-RoGO system can also be applied to gait rehabilitation in incomplete motor SCI. It can be hypothesized that coupling the behavioral activation of the supraspinal gait areas (via the BCI) and spinal cord gait central pattern generators (feedback driving via the RoGO) may provide a unique form of Hebbian learning. This could potentially improve neurological outcomes after incomplete SCI beyond those of standard gait therapy. 

\section*{Acknowledgment}

This project was funded by the Long Beach Veterans Affairs Southern California Institute for Research and Education (SCIRE) Small Projects Grant, the Long Beach Veterans Affairs Advanced Research Fellowship Grant, the American Brain Foundation, the National Institutes of Health (Grant UL1 TR000153), and the National Science Foundation (Award: 1160200).

\ifCLASSOPTIONcaptionsoff
  \newpage
\fi



%

\bibliographystyle{unsrt}
\bibliography{IEEEabrv,BCIRoGO}


%




\end{document}